\documentclass[11 pt,leqno]{article} \usepackage{latexsym}
\usepackage{amssymb,amsbsy,amsmath,amsfonts,amscd}
\usepackage{vmargin}

\usepackage{graphicx}
\usepackage[english]{babel}
\usepackage{amsmath}
\usepackage{alltt}

\newlength\jataille
\newcommand{\figgauche}[5]%
{\jataille=\textwidth\advance\jataille by -#1
\advance\jataille by -.5cm
\begin{figure}[!h]
\begin{minipage}[c]{#1}
   \includegraphics[width=#1]{#2}
\caption{ #3}
\label {#4}
\end{minipage}\hfill
\begin{minipage}[c]{\jataille}
   \footnotesize #5 \normalsize
\end{minipage}
\end{figure}}

\usepackage{epsfig}

\newtheorem{theorem}{Theorem}[section]

 \newcommand{\N}{\ifmmode{{\rm I} \hskip -2pt {\rm R}}
    \else{\hbox{$I\hskip -2pt N$}}\fi}

 \newcommand{\R}{\ifmmode{{\rm I} \hskip -2pt {\rm R}}
    \else{\hbox{$I\hskip -2pt R$}}\fi}

\newcommand{\BEQ} {\begin{equation} }
\newcommand{\EEQ} {\end{equation} }

\makeatletter
\@addtoreset{equation}{section}

\begin{document}

\title{Stability of some turbulent vertical models for the ocean mixing boundary layer}

\author{{\small A -C.\ BENNIS}\thanks{%
IRMAR, Universit\'{e} de Rennes 1, Campus de Beaulieu, 35042 Rennes Cedex,
France}\ ,\thinspace {\small T.\
CHAC\'{O}N\ REBOLLO, M.\ GOMEZ MARMOL} \thanks{%
Departamento de Ecuaciones Diferenciales y An\'{a}lisis Numerico,
Universidad de Sevilla.\ C/Tarfia, s/n.41080, Sevilla, Spain}\ ,\thinspace
{\small R.\ LEWANDOWSKI} \thanks{%
IRMAR, Universit\'{e} de Rennes 1, Campus de Beaulieu, 35042 Rennes Cedex,
France} ,}
\date{}
\maketitle

\begin{abstract} We consider four turbulent models to simulate the boundary mixing layer of the ocean. We show the existence of solutions to these models in the steady-state case then we study the mathematical stability of these solutions. 
\end{abstract} 

\medskip


\section{Introduction}

The presence of an homogeneous layer near the surface of the ocean
has been observed since a long time.\ The so called ''mixed layer''
presents almost constant profiles of temperature and salinity (or equivalently the density).\ The
bottom of the mixed layer corresponds either to the top of the
thermocline, zone of large gradients of temperature, or to the top
of the zone where haline stratification is observed \cite{Vi98}.
Some attempts to describe this phenomenon can be found for example
in Defant \cite{De36} or Lewandowski \cite{Lew97}. The effect of
the wind-stress acting on the sea-surface was then considered to be
the main forcing of this boundary layer.\ Observations in situ were
completed by laboratory experiments \cite{Dea69} and more recently
by numerical modelizations of the mixed layer.

\medskip

In this note, we consider four turbulent models to describe this homogeneous boundary layer. The first one is the Pacanowski-Philander model, and two of these models are new models. They aim to compute the velocity and the water density of a water column, are one space dimensional  and the eddy viscosities depend on the Richardson number. For those model, we show the existence of a steady-state solution and we analyse the mathematical linear stability of these steady state solution, showing that only one of these model, the one we introduce in this note (model labelized as $R-2-2-4$ below), has a unique staedy state solution with a large range of stabilty. Moreover, in \cite{BCGL07} we have used these models to simulate the warm pool at the equator.  Numerical results confirm that 
$R-2-2-4$ is the most accurate parametrization. 

\section{The equations} 

We denote by $(u,v)$ the horizontal water velocity and $\rho$ its density. Since the numerical simulation performed in $ \cite{BCGL07}$ concerns the equator zone, we do not take the Coriolis force into account. The closure equations are:

\begin{equation} \label{EQS}
\left\{
\begin{array}{l}
\dfrac{\partial u}{\partial t}-\dfrac{\partial }{\partial z}\left( \nu _{1}%
\dfrac{\partial u}{\partial z}\right) =0, \quad

\dfrac{\partial v}{\partial t}-\dfrac{\partial }{\partial z}\left( \nu _{1}%
\dfrac{\partial v}{\partial z}\right) =0, \\\\

\dfrac{\partial \rho }{\partial t}-\dfrac{\partial }{\partial z}\left( \nu
_{2}\dfrac{\partial \rho }{\partial z}\right) =0,\text{ for }t\geqslant 0%
\text{ and }-h\leqslant z\leqslant 0, \\\\

u=u_{b},\,\,v=v_{b},\,\,\rho =\rho _{b}\text{ \ at the depth }z=-h, \\\\ \displaystyle 
\nu _{1}\dfrac{\partial u}{\partial z}=\frac{\rho_{a}}{\rho_{0}}V_{x},\,\,\nu _{1}\dfrac{\partial v}{\partial z}=\frac{\rho_{a}}{\rho_{0}}V_{y},\,\, \nu _{2}\dfrac{\partial \rho }{%
\partial z}=Q\text{ \ at the surface }z=0, \\\\
u=u_{0},\,\,v=v_{0},\,\,\rho =\rho _{0}\text{ \ at initial time }t=0.
\end{array}
\right.
\end{equation}

\medskip

In system $(\ref{EQS})$, the coefficients $\nu _{1}$ and $\nu _{2}$ are the vertical eddy viscosity
and diffusivity coefficients and will be expressed as functions of the
Richardson number $R$ defined as
$$ \displaystyle
R=\dfrac{-\dfrac{g}{\rho_{0}}\cdot\dfrac{\partial \rho }{\partial z%
}}{\left( \dfrac{\partial u}{\partial z}\right) ^{2}+\left( \dfrac{\partial v}{\partial z}\right) ^{2}} $$
where $g$ is the gravitational acceleration and $\rho_{0}$ a reference
density ($\rho_{0}\simeq 1025\hskip 2 pt kg.m^{-3})$.

\medskip

The constant $h$ denotes the thickness of the studied layer
that must contain the mixing layer.\ Therefore the circulation for $z<-h$,
under the boundary layer, is supposed to be known, either by observations or
by a deep circulation numerical model. This justifies the choice of
Dirichlet boundary conditions at $z=-h$, $u_{b}$, $v_{b}$ and $\rho _{b}$ being the
values of horizontal velocity and density in the layer located below the
mixed layer.\ The air-sea interactions are represented by the fluxes at the
sea-surface : $V_{x}$ and $V_{y}$ are respectively the forcing exerced by the zonal wind-stress and the meridional wind-stress and
$Q$ represents the thermodynamical fluxes, heating or cooling,
precipitations or evaporation.
We have $V_{x}=C_{D}\left|
u^{a}\right| ^{2}$ and $V_{y}=C_{D}\left|
v^{a}\right| ^{2}$, where $U^{a}=(u_{a},v_{a})$ is the air velocity and $C_{D}$ a friction
coefficient.

\medskip
We study hereafter four different formulations for the eddy coefficients $%
\nu _{i}=f_{i}\left( R\right) $, labeled as $"R-2-i"$ and/or  $"R-2-i-j"$ . In all models, 
$f_{1}\left( R\right) =\alpha _{1}+\dfrac{\beta _{1}}{\left( 1+5R\right) ^{2}}$, except in model $R-2-3$ below: 
\begin{equation}  \label{MODELES} \begin{array} {l} R-2-1-3: \quad \displaystyle 
f_{2}\left( R\right) =\alpha _{2}+\dfrac{f_{1}\left( R\right) }{1+5R%
}=\alpha _{2}\text{+}\frac{\alpha _{1}}{1+5R}\text{+}\frac{\beta _{1}}{%
\left( 1+5R\right) ^{3}}. \\ R-2-3 \quad f_{1}\left( R\right) =\alpha _{1}+\dfrac{\beta _{1}}{\left( 1+10R\right) ^{2}}, \quad
f_{2}\left( R\right) =\alpha _{2}+\dfrac{\beta _{2}}{\left(
1+10R\right) ^{3}} \\ R-2-2-4: \quad \displaystyle
f_{2}\left( R\right) =\alpha _{2}+\dfrac{f_{1}\left( R\right) }{%
(1+5R)^{2}}=\alpha _{2}+\frac{\alpha _{1}}{(1+5R)^{2}}+\frac{\beta _{1}}{%
\left( 1+5R\right) ^{4}},\\ R-2-2 \quad
f_{2}\left( R\right) =\alpha _{2}+\dfrac{\beta _{2}}{\left(
1+5R\right) ^{2}}, \end{array}
\end{equation}
Formulation $R-2-1-3$ corresponds to the modelization of the vertical mixing
proposed by Pacanowski and Philander \cite{Pa81}. The
coefficients $\alpha _{1},\beta _{1}$and $\alpha _{2}$ have the following values:
$\ \ \ \alpha _{1}=1.10^{-4},\,\,\beta _{1}=1.10^{-2},\,\,\ \alpha
_{2}=1.10^{-5}($units:$m^{2}s^{-1}).$This formulation has been used in the
OPA code developed in Paris 6 University \cite{Ma97}
 with coefficients $\alpha _{1}=1.10^{-6},\,\,\beta
_{1}=1.10^{-2},\,\,\alpha _{2}=1.10^{-7}\left( \text{units:\thinspace }%
m^{2}s^{-1}\right) $.The selection criterion for the coefficients appearing
in these formulas was the best agreement of numerical results with
observations carried out in different tropical areas.
Formulation $R-2-3$ has been proposed by Gent \cite{Gen91}. Formulations 
$R-2-2-4$ and $R-2-2$ are new as far as we know. Notice that 
models $R-2-1-3$ and $R-2-3$ are no more physically valid respectively for 
$R \in (-3.13, -0.2)$ and $R \in (-2.25, -0.1)$ since
the coefficient $\nu _{2}$ becomes negative. 

\subsection{Steady-state solutions}

Steady-state solutions to system $(\ref{EQS})$ satisfy

\begin{equation} \label {STS} 
\dfrac{\partial }{\partial z}\left( f_{1}\left( R\right) \dfrac{\partial u}{%
\partial z}\right) =0, \quad 
\dfrac{\partial }{\partial z}\left( f_{1}\left( R\right) \dfrac{\partial v}{%
\partial z}\right) =0,\quad 
\dfrac{\partial }{\partial z}\left( f_{2}\left( R\right) \dfrac{\partial
\rho }{\partial z}\right) =0.
\end{equation}

\begin{theorem} System $(\ref{STS})$ has at leat one smooth solution on $[0,-h]$ for each model in 
$(\ref {MODELES})$. In case of $R-2-2-4$ the solution is unique. 
\end{theorem} 

{\bf Proof.} 
Integrating $(\ref{STS})$ with respect to $z,$ yields
\begin{equation}\label{int}
\displaystyle f_{1}\left( R\right) \dfrac{\partial u}{\partial z} 
=\frac{V_{x}\rho_{a}}{\rho_{0}}, \quad 
\displaystyle f_{1}\left( R\right) \dfrac{\partial v}{\partial z}
=\frac{V_{y}\rho_{a}}{\rho_{0}},\quad 
\displaystyle f_{2}\left( R\right) \dfrac{\partial \rho }{\partial z} = Q.
\end{equation}

and since $R=\dfrac{-\dfrac{g}{\rho_{0}}\cdot\dfrac{\partial \rho }{\partial z%
}}{\left( \dfrac{\partial u}{\partial z}\right) ^{2}+\left( \dfrac{\partial v}{\partial z}\right) ^{2}}$ \ we deduce from $(\ref{int})$ 
that

$
\displaystyle R=-\dfrac{g Q \rho_{0}}{\rho_{a}^{2}(V_{x}^{2}+V_{y}^{2})}\cdot\dfrac{\left( f_{1}\left(
R\right) \right) ^{2}}{f_{2}\left( R\right) }
$, which yields
\begin{equation}\label{pf}
\dfrac{\left( f_{1}\left( R\right) \right) ^{2}}{f_{2}\left( R\right) }=-%
\frac{\rho_{a}^{2}(V_{x}^{2}+V_{y}^{2})}{g Q \rho_{0}}R
\end{equation}
which is a fixed point equation for $R$.
\medskip
Any solution $R$ to equation $(\ref{pf})$ yields a 
Richardson number $R^{e}$ corres
ponding to the fluxes $V_{x}$,$V_{y}$ and $Q$ and not on $z$ as $\nu _{1}$ and $%
\nu _{2}$ are independent on the depth variable $z$ as well asq the turbulent viscosities. The Richardson number $%
R^{e}$ being known, steady-state profiles for velocity and density are
obtained by integrating $(\ref{int})$ with respect to $z$, taking into
account the boundary conditions at $z=-h$:
\begin{equation} \begin{array} {l} 
u^{e}\left( z\right) =u_{b}+\dfrac{V_{x}\rho_{a}}{\rho_{0}f_{1}\left( R^{e}\right) }\,\left(
z+h\right) ,\quad 
v^{e}\left( z\right) =v_{b}+\dfrac{V_{y}\rho_{a}}{\rho_{0}f_{1}\left( R^{e}\right) }\,\left(
z+h\right) ,\\
\rho ^{e}(z)=\rho _{b}+\dfrac{Q}{f_{2}\left( R^{e}\right) }\,\left(
z+h\right) . \end{array}
\end{equation}
It remains to analyse the existence of solutions of equation $(\ref{pf})$. These solutions can be interpreted as the intersection of the
curves $k\left( R\right) =\dfrac{\left( f_{1}\left( R\right) \right) ^{2}}{%
f_{2}\left( R\right) }$ and $h\left( R\right) =CR$ with $\displaystyle C=-\frac{\rho_{a}^{2}(V_{x}^{2}+V_{y}^{2})}{g Q \rho_{0}}%
$. The existence and the number of solutions are
controlled by the constant $C$ and then by the parameter $\dfrac{V^{2}}{Q}$, 
$V^2 = V_x^2 + V_y^2$, 
depending only on the surface fluxes.
The graph of function $k$ and $h$ for $Q<0$ and $Q>0$ is plotted on Figures 1 and 2 below when $f_{1}$ and $%
f_{2}$ in case of R-2-2-4 and $R-2-2$.

\begin{center}
\begin{tabular}{ccc}
\includegraphics[scale=0.45]{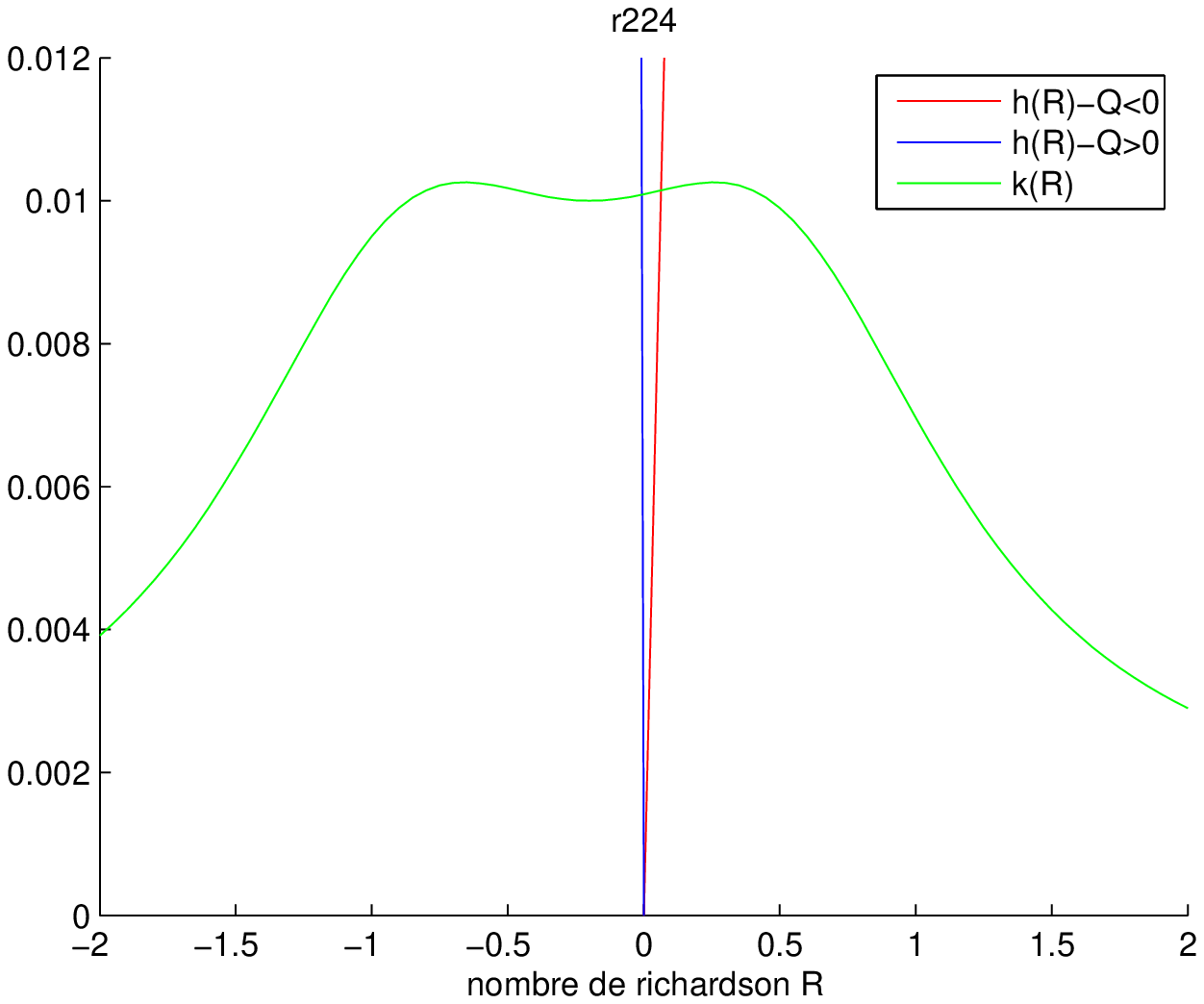}  & & 
\includegraphics[scale=0.45]{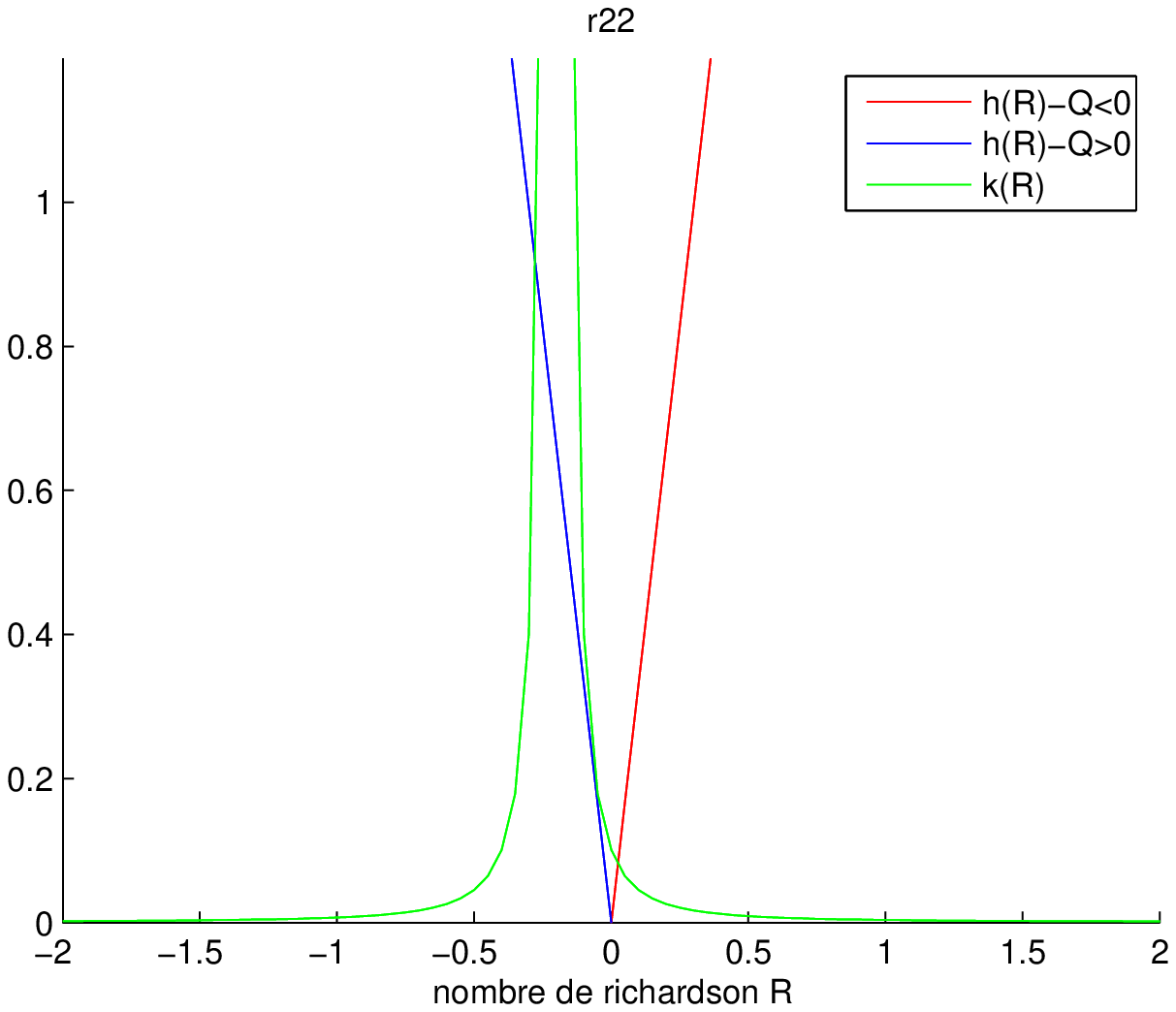} \\
\begin{tabular}{c}
\\
Formulation R-2-2-4\protect\medskip \\
\textbf{Figure A}
\end{tabular}
 & &
\begin{tabular}{c}
\\
Formulation R-2-2\protect\medskip \\
\textbf{Figure B}
\end{tabular}
\end{tabular}
\end{center}
The qualitative behaviour obtained with formulation R-2-3 and R-2-1-3 is the same as R-2-2. The intersection of $k(R)$ and $h(R)=CR$ consists in one
point for $Q<0$ and several points for $Q>0$. The number of points depends to the values of surface fluxes.

The graphs obtained for the R-2-2-4 modelization (Figure A) and its
simplified version R-2-2 (Figure B) are very different.\ It is obvious in
Figure A that any straight line $h\left( R\right) =CR$ meets $k$ at only one
point for $Q>0$ and $Q<0$.\ Therefore it exists one unique equilibrium Richardson number $R^{e}$
whatever the values of the surface fluxes $V_x$, $V_y$ and $Q$. In the case of the other models, we get several solutions. The proof is finished. Notice that in \cite{BCGL07} we show that the most accurate model is $R-2-2-4$ from the physical and numerical viewpoint.

\subsection{Linear stability of the equilibrium solutions}

In this section we analyse the time evolution of a small perturbation of one
of the equilibrium states $\left( u^{e},v^{e},\rho ^{e}\right) $ described in the
previous section.

At initial time $t=0$ we set
$
\left( u_{0},v_{0},\rho _{0}\right) =\left( u^{e},v^{e},\rho ^{e}\right) +\left(
u_{0}^{\prime },v_{0}^{\prime},\rho _{0}^{\prime }\right)
$
and we denote by
\begin{equation*}
\left( u,v,\rho \right) =\left( u^{e},v^{e},\rho ^{e}\right) +\left( u^{\prime
},v^{\prime},\rho ^{\prime }\right)
\end{equation*}
the solution of equations $(\ref{EQS})$ at time $t$ where  $\left( u^{e},v^{e},\rho ^{e}\right) $  are solution to the steady-state system $(\ref{STS})$, 
and $\nu _{1}^{e}=f_{1}\left( R^{e}\right) $ and $\nu _{2}^{e}=f_{2}\left(
R^{e}\right) $ are two positive constants.

Introducing the new variables $\psi =\dfrac{\partial \rho }{\partial z}$,
$\theta =\dfrac{\partial u}{\partial z}$ and $\beta =\dfrac{\partial v}{\partial z}$, the Richardson number can be
expressed as $$R=-\dfrac{g}{\rho_{0}}\,\dfrac{\psi }{(\theta ^{2}+\beta^{2})}=R\left(
\theta,\beta,\psi \right) $$

Applying the Taylor formula, we get
$$ \begin{array}{l}
{\cal{F}}=
\left( \theta -\theta ^{e}\right) \,\dfrac{\partial \nu _{1}}{%
\partial \theta }\,\left( \theta ^{e},\beta^{e},\psi ^{e}\right)
+\left( \beta -\beta ^{e}\right) \,\dfrac{\partial \nu _{1}}{%
\partial \beta }\,\left( \theta^{e},\beta ^{e},\psi ^{e}\right)
+\left( \psi -\psi
^{e}\right) \,\dfrac{\partial \nu _{1}}{\partial \psi }\,\left( \theta
^{e},\beta^{e},\psi ^{e}\right) +\cdots \, \\
{\cal{G}}=
\left( \theta -\theta ^{e}\right) \,\dfrac{\partial \nu _{2}}{%
\partial \theta }\,\left( \theta ^{e},\beta^{e},\psi ^{e}\right)
+\left( \beta -\beta ^{e}\right) \,\dfrac{\partial \nu _{2}}{%
\partial \beta }\,\left( \theta^{e},\beta ^{e},\psi ^{e}\right)
+\left( \psi -\psi
^{e}\right) \,\dfrac{\partial \nu _{2}}{\partial \psi }\,\left( \theta
^{e},\beta^{e},\psi ^{e}\right) +\cdots \, \end{array} 
$$
We set for
$k=1,2$ : ${\cal{F}}=\nu _{1}\left( \theta,\beta,\psi \right)-\nu _{1}\left( \theta ^{e},\beta^{e},\psi
^{e}\right)$, ${\cal{G}}=\nu _{2}\left( \theta,\beta,\psi \right)-\nu _{2}\left( \theta ^{e},\beta^{e},\psi
^{e}\right)$
and
$\nu _{k}^{e} =\nu _{k}\left( \theta ^{e},\beta^{e},\psi ^{e}\right)$, $\theta
^{^{\prime }}=\theta -\theta ^{e},\,\beta ^{\prime }=\beta -\beta ^{e}$, $\psi ^{\prime }=\psi -\psi ^{e}$, 
$$\left( \dfrac{\partial \nu _{k}}{\partial \theta }\right) ^{e}=\dfrac{%
\partial \nu _{k}}{\partial \theta }\left( \theta ^{e},\beta^{e},\psi ^{e}\right)
,\,\left( \dfrac{\partial \nu _{k}}{\partial \beta }\right) ^{e} =\dfrac{%
\partial \nu _{k}}{\partial \beta }\left( \theta ^{e},\beta^{e},\psi ^{e}\right)
,\,\,\,\left( \dfrac{\partial \nu _{k}}{\partial \psi }\right) ^{e}=\dfrac{%
\partial \nu _{k}}{\partial \psi }\left( \theta ^{e},\beta^{e},\psi ^{e}\right) .
$$
The equations satisfied by the perturbation $\left( u^{\prime },v^{\prime },\rho
^{\prime }\right) $ are deduced from equations $(\ref{EQS})$: 
\begin{equation}
\left\{
\begin{array}{l}
\dfrac{\partial u^{\prime }}{\partial t}-\dfrac{\partial }{\partial z}\left(
\nu _{1}\left( \theta,\beta ,\psi \right) \,\left( \theta ^{e}+\theta ^{\prime
}\right) \right) =0,\quad 
\dfrac{\partial v^{\prime }}{\partial t}-\dfrac{\partial }{\partial z}\left(
\nu _{1}\left( \theta,\beta ,\psi \right) \,\left( \beta ^{e}+\beta ^{\prime
}\right) \right) =0,\\
\dfrac{\partial \rho ^{\prime }}{\partial t}-\dfrac{\partial }{\partial z}%
\left( \nu _{2}\left( \theta,\beta ,\psi \right) \,\left( \psi ^{e}+\psi ^{\prime
}\right) \right) =0.
\end{array}
\right.
\end{equation}
We now replace $\nu _{1}$ and $\nu _{2}$ by expresions deuced from the Taylor's development  and retain only the first order terms.\ The approximated
equations for $\left( u^{\prime },v^{\prime },\rho ^{\prime }\right) $ then are
\begin{equation}\label{15}
\left\{
\begin{array}{c}
\dfrac{\partial u^{\prime }}{\partial t}-\dfrac{\partial }{\partial z}\left(
\left( \nu _{1}^{e}+\theta ^{e}\left( \dfrac{\partial \nu _{1}}{\partial
\theta }\right) ^{e}\right) \theta ^{\prime }\right)-\dfrac{\partial }{%
\partial z}\left( \theta ^{e}\left( \dfrac{\partial \nu _{1}}{\partial \beta }%
\right) ^{e}\beta ^{\prime }\right) -\dfrac{\partial }{%
\partial z}\left( \theta ^{e}\left( \dfrac{\partial \nu _{1}}{\partial \psi }%
\right) ^{e}\psi ^{\prime }\right) =0,\medskip \\

\dfrac{\partial v^{\prime }}{\partial t}-\dfrac{\partial }{%
\partial z}\left( \beta^{e}\left( \dfrac{\partial \nu _{1}}{\partial \theta }%
\right) ^{e}\theta^{\prime }\right)-\dfrac{\partial }{\partial z}\left(
\left( \nu _{1}^{e}+\beta^{e}\left( \dfrac{\partial \nu _{1}}{\partial
\beta }\right) ^{e}\right) \beta ^{\prime }\right)-\dfrac{\partial }{%
\partial z}\left( \beta^{e}\left( \dfrac{\partial \nu _{1}}{\partial \psi }%
\right) ^{e}\psi ^{\prime }\right) =0,\medskip \\

\dfrac{\partial \rho ^{\prime }}{\partial t}-\dfrac{\partial }{\partial z}%
\left( \psi ^{e}\left( \dfrac{\partial \nu _{2}}{\partial \theta }\right)
^{e}\theta ^{\prime }\right)-\dfrac{\partial }{\partial z}%
\left( \psi ^{e}\left( \dfrac{\partial \nu _{2}}{\partial \beta }\right)
^{e}\beta^{\prime }\right) -\dfrac{\partial }{\partial z}\left( \left( \nu
_{2}^{e}+\psi ^{e}\left( \dfrac{\partial \nu _{2}}{\partial \psi }\right)
^{e}\right) \psi ^{\prime }\right) =0.
\end{array}
\right.
\end{equation}

We set
\begin{equation*}
A=\left(
\begin{array}{lll}
\nu _{1}^{e}+\theta ^{e}\left( \dfrac{\partial \nu _{1}}{\partial \theta }%
\right) ^{e} & \theta ^{e}\left( \dfrac{\partial \nu _{1}}{\partial \beta }%
\right) ^{e}&\theta ^{e}\left( \dfrac{\partial \nu _{1}}{\partial \psi }%
\right) ^{e} \\\\

\beta ^{e}\left( \dfrac{\partial \nu _{1}}{\partial \theta }%
\right) ^{e}&\nu _{1}^{e}+\beta ^{e}\left( \dfrac{\partial \nu _{1}}{\partial \beta }%
\right) ^{e} &\beta ^{e}\left( \dfrac{\partial \nu _{1}}{\partial \psi }%
\right) ^{e} \\\\

\psi ^{e}\left( \dfrac{\partial \nu _{2}}{\partial \theta }\right) ^{e}&\psi ^{e}\left( \dfrac{\partial \nu _{2}}{\partial \beta }\right) ^{e}
& \nu _{2}^{e}+\psi ^{e}\left( \dfrac{\partial \nu _{2}}{\partial
\psi }\right) ^{e}
\end{array}
\right) ,\,\ \ V=\left(
\begin{array}{c}
u^{\prime } \\
v^{\prime } \\
\rho ^{\prime }
\end{array}
\right) ,
\end{equation*}
Equations $(\ref{15})$ can be written
\begin{equation} \label{VEQ}
\dfrac{\partial V}{\partial t}-\dfrac{\partial }{\partial z}\left( A\dfrac{%
\partial V}{\partial z}\right) =\dfrac{\partial V}{\partial t}-A\dfrac{%
\partial ^{2}V}{\partial z^{2}}=0.
\end{equation}

Let $\left( \lambda _{1},\lambda _{2},\lambda _{3}\right) $ be the eigenvalues of
matrix $A$. Assuming the eigenvalues dinstincts,matrix $A$ is equal to $P^{-1}DP$, where $D$ is diagonal,
and such that $d_{11}=\lambda _{1}$, $d_{22}=\lambda _{2}$ and $d_{33}=\lambda _{3}$. Set now $%
W=PV. $ The vector $W$ verifies the system $\dfrac{\partial W}{\partial t}%
-D\,\dfrac{\partial ^{2}W}{\partial z^{2}}=0,\,\,\,\ $i.e.

\begin{equation}
\dfrac{\partial w_{1}}{\partial t}-\lambda _{1}\dfrac{\partial ^{2}w_{1}}{%
\partial z^{2}}=0,\quad
\dfrac{\partial w_{2}}{\partial t}-\lambda _{2}\dfrac{\partial ^{2}w_{2}}{%
\partial z^{2}}=0,\quad
\dfrac{\partial w_{3}}{\partial t}-\lambda _{3}\dfrac{\partial ^{2}w_{3}}{%
\partial z^{2}}=0.
\end{equation}

Stability of the equilibrium solution $\left( u^{e},v^{e},\rho ^{e}\right) $
means that any perturbation $\left( u_{0}^{\prime },v_{0}^{\prime },\rho _{0}^{\prime
}\right) $ imposed at initial time $t=0$ is damped as $t\rightarrow \infty .$
This is verified if the eigenvalues $\lambda _{1},\lambda _{2},\lambda_{3}$  are such
that ${Re}\left( \lambda _{1}\right) >0$, ${Re}\left( \lambda
_{2}\right) >0$ and ${Re}\left( \lambda
_{3}\right) >0$. 
These three conditions are  equivalence to 
$\det \left( A\right) >0$, 
$\text{tr}\left( A\right) >0$ and $\text{tr}\left( Adj(A)\right) >0$.
From these conditions, we build the graph below (see figure 1), obtained thanks an analytical  computation (we skip the technical details here):

\begin{center}

\figgauche{9cm}{graph}{Numerical stability}{geom}{ The results are summarized in Figure 1.
The circle zone represents a zone where the solution is physically not valid. It is the case for the
R-2-3 and R-2-1-3 formulation. The rectangular zone is a unstability zone. All formulations have a
unstability zone.  Nevertheles, one observes that for each model,  mathematical stability holds for non negative $R$. 

 } 
\end{center}

\section{Conclusion}  All the models  have a steady-state solution, unique in the case of R-2-2-4. Each one is linearly stable for non negative $R$, which corresponds to physical stability. All these models present a linear unstable zone, located in a region where $R$ is non positive. They all presents a linear stability zone for some non positive values of $R$, situation that can arise in real situation, as reported in  $\cite{BCGL07}$ (physical unstability). All these models have been tested in $\cite{BCGL07}$. The simulation confirms the existence of stable linear steady-state solutions and the ability of these models to describe a boundary mixing layer. However, the numerical study in  $\cite{BCGL07}$ confirms that $R-2-2-4$ yields better numerical results.


\bibliographystyle{siam} \bibliography{Bib}

\end{document}